# Elastic Modulus Versus Cell Packing Density in MDCK Epithelial Monolayers

Steven J. Chisolm[1*], Emily Guo[2], Vignesh Subramaniam[1], Kyle D. Schulze[2], Thomas E. Angelini[1,3,4]

[1]Department of Mechanical and Aerospace Engineering, University of Florida, Gainesville, FL, 32605.

[2] Department of Mechanical Engineering, Auburn University, Auburn, AL, 36849.

[3]Department of Materials Science and Engineering, University of Florida, Gainesville, FL, 32605.

[4]J. Crayton Pruitt Family Department of Biomedical Engineering, University of Florida, Gainesville, FL, 32605.

*Corresponding author. Email address: stevenchisolm@ufl.edu

The elastic moduli of tissues are connected to their states of health and function. The epithelial monolayer is a simple, minimal, tissue model that is often used to gain understanding of mechanical behavior at the cellular or multi-cellular scale. Here we investigate how the elastic modulus of Madin Darby Canine Kidney (MDCK) cells depends on their packing density. Rather than measuring elasticity at the sub-cellular scale with local probes, we characterize the monolayer at the multi-cellular scale, as one would a thin slab of elastic material. We use a micro-indentation system to apply gentle forces to the apical side of MDCK monolayers, applying a normal force to approximately 100 cells in each experiment. In low-density confluent monolayers, we find that the elastic modulus decreases with increasing cell density. At high densities, the modulus appears to plateau. This finding will help guide our understanding of known collective behaviors in epithelial monolayers and other tissues where variations in cell packing density are correlated with cell motion.

# 1. Introduction

The elasticity of living cells is intimately coupled to their function across different states of tissue health, development, and disease. For example, in the lung and the breast, cancerous tissue often exhibits a significantly lower elastic modulus than healthy functional tissue (Cross et al. 2020). During body axis elongation in zebrafish embryos, developing tissue exhibits spatially varying material properties, where elasticity at one end preserves supracellular structure while dominantly fluid-like behavior at the other end enables cell rearrangements and flow during extension and growth (Mongera et al. 2018). Gradients in cell size, number density, and cell elasticity have been found to correlate to the aggressiveness of tumor invasion into the surrounding micro-environment (Han et al. 2020). To guide our interpretation of these complex phenomena, simple *in vitro* cellular models are often employed for highly controlled experimentation and the testing of theories. For example, much of our current understanding of the relationship between cell packing density and patterns of motion came from studies of monolayers as model tissues (Angelini et al. 2010, Angelini et al. 2011, Bi et al. 2015, Bi et al. 2011, Chisolm et al. 2024, Nnetu et al. 2012, Tambe et al. 2011, Trepat et al. 2009). While a great deal of focus has been given to motion and mechanical forces in monolayers, much less is known about the relationship between packing density and elastic modulus. The elastic modulus of individual cells in monolayers have been measured using methods such as magnetic twisting cytometry (MTC), in which a rotating magnetic field applies torsion to ferromagnetic beads attached to the cytoskeleton (Barry et al. 2015, Overby et al. 2014, Trepat et al. 2006). Similarly, magnetic tweezers (MT) have been used to determine cytoskeletal material properties by driving magnetic beads contained inside the cells within monolayers (Huang et al. 2005). By using atomic force microscopy (AFM), it was found that the elastic modulus of human umbilical vein endothelial cells (HUVECs) increased with increasing spread area, mediated by cell-cell cohesion (Stroka and Aranda-Espinoza 2011). While microscopic techniques like MTC, MT, and AFM have enabled measurements of elasticity of isolated cells and cells in monolayers, it is not clear how these measurements at the sub-cellular scale are related to the elasticity of monolayers or tissues when tested as macro-scale materials. Few investigations of the material properties of monolayers at multi-cellular scales have been performed (Schulze et al. 2017), and a systematic investigation of monolayer elasticity at different cell packing densities remains to be carried out. Making the connection between monolayer elasticity at multi-cellular scales and cell packing density would represent a valuable step toward understanding the hierarchical relationships that determine tissue mechanics in health and disease.

In this manuscript we investigate the relationship between the elastic modulus and cell number density in a model epithelial monolayer. We perform indentation measurements on confluent islands of Madin Darby Canine Kidney cells, collecting force-indentation curves on layers prepared at different seeding densities, indenting approximately 100 cells in each measurement. We find the force-indentation profiles exhibit a strong dependence on cell density, and we employ a simple contact mechanics model to determine their elastic moduli. We find that the elasticity of MDCK monolayers decreases with increasing cell number density within the lower range of densities, plateauing in the higher range. Given the well-established connection between collective motion in monolayers and cell density, our findings indicate that cell density fluctuations may be accompanied by spatiotemporal patterns of varying elasticity that mediate cell-cell coupling and the patterns in collective motion that emerge.

## 2. Results

### 2.1 Micro-indentation tests on cell monolayers

To create epithelial monolayers for indentation tests, we seed cells at various number densities onto collagen-coated glass-bottomed culture dishes. We vary the number density by depositing 2 μL droplets of liquid growth media containing MDCK cells dispersed at different concentrations. Once the cells have adhered, we gently add 2 mL of culture media to the dishes and incubate overnight. To enable cell density to be determined, we dye the confluent islands with 5-chloromethylfluorescein diacetate (CMFDA). Indentation experiments are performed on an inverted microscope equipped with a heated stage that maintains the culture dishes at 37°C. The indenting probe-tip is a 1 mm diameter hemisphere made from borosilicate glass coated in F-127 Pluronic which prevents adhesion to the cells. The probe tip is manually positioned approximately 2 μm above the apical surface of the monolayer and translated downward at a rate of 1 μm/s, indenting the monolayers by approximately 3 μm at the tip's apex. The indentation force and indentation depth are recorded at a rate of 5000 Hz while fluorescence images are simultaneously recorded (see Methods for instrument details). We use these images to assess the condition of the monolayers before, during, and after indentation, ensuring the process causes no damage (Fig. 1).

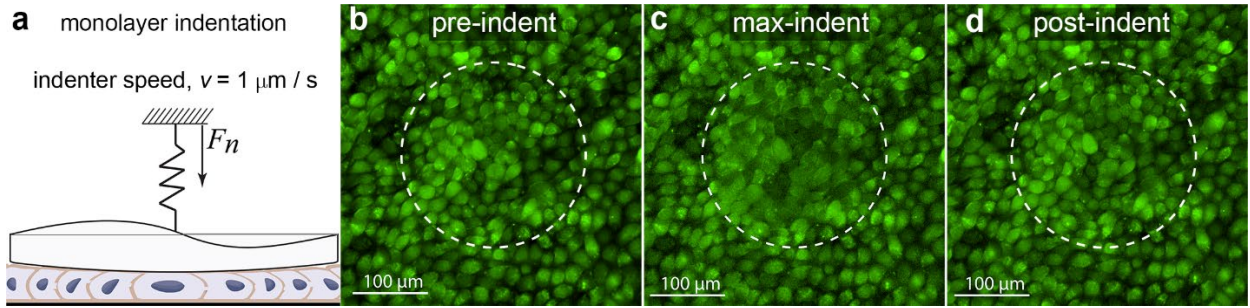

**Figure 1.** Monolayer micro-indentation experiments. (a) A micro-indenter is mounted above an inverted microscope. The hemispherical indentation probe moves at a rate of 1μm/s as it indents the MDCK monolayer. The instrument records the normal force and the indentation depth throughout the experiment. (b) Images of the monolayer collected prior to indentation, (c) at the maximum indentation depth, (d) and a short time after indentation.

To determine whether the elastic modulus of monolayers depends on cell density, we analyze each force-displacement (*F-d*) dataset and we perform image analysis to measure the cell number density of the corresponding monolayer. Briefly, cell density is measured using a machine-learning based image segmentation software, Cellpose 2.0 (Stringer et al. 2021), employing the pre-trained Cyto2 model. The number of identified cells is divided by the area of the field of view to determine the cell number density. Representative data collected on monolayers having different number densities demonstrate that the indentation force, *F*, rises more rapidly with increasing indentation depth, *d*, than for monolayers at lower densities (Fig. 2a).

To determine an elastic modulus from each measured *F-d* dataset, we restrict our analysis to the smallest range of indentation depths over which a sufficiently large dynamic range of measured forces occurs across all samples. Working in the thin-slab limit, we estimate the estimate the average normal strain in the monolayer to be given by $\langle \gamma \rangle = \langle d \rangle / h$, where *h* is the monolayer thickness and the angle brackets indicate an average over the contacting surface area. It can be shown that that $\langle d \rangle = d_{max}/2$ for a spherical indenter where $d_{max}$ is the indentation depth at the apex of the indenter. Thus, by limiting $d_{max}$ to 2.4 μm, the average strain in the monolayer at the high end of the indent is 0.2 for a representative monolayer having a thickness of *h* = 6 μm. Our analysis below indicates that the monolayers remain in the linear elastic regime during indentation measurements when employing this strategy.

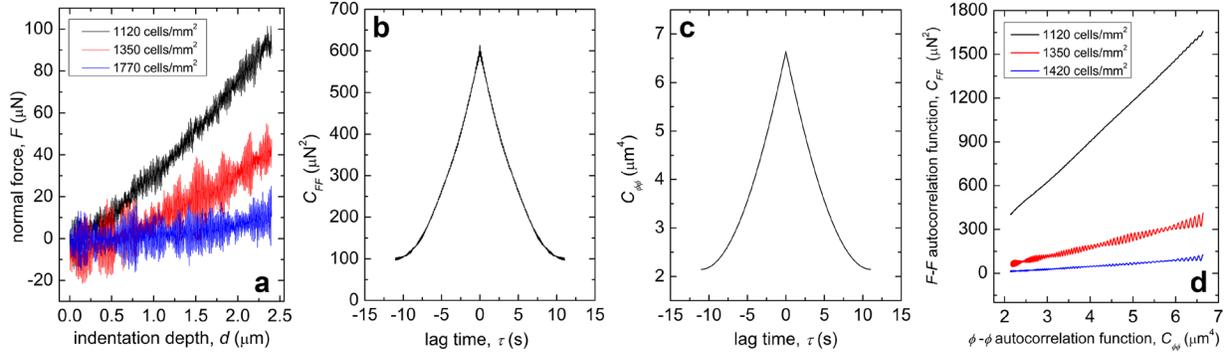

**Figure 2.** Indentation and autocorrelation analysis. (a) Force-displacement curves of indented monolayers exhibit high levels of noise. (b) Plots of the force autocorrelation function, $C_{FF}$, demonstrate how statistical averaging reduces the contribution of random noise in our analysis method. (c) The autocorrelation function of $\phi = d^2$, $C_{\phi\phi}$, visibly resembles $C_{FF}$ and exhibits even less noise. (d) Plots of $C_{FF}$ vs $C_{\phi\phi}$ for monolayers having different densities exhibit linear trends. The indentation modulus of each monolayer is determined linear regression analysis of these plots.

## 2.2 Correlation analysis of noisy force-indentation curves

We find that that the noise amplitude in $F$-$d$ measurements is approximately ±10 μN, which could potentially reduce our confidence in determining the elastic modulus of many monolayers, especially those in the higher density range. To systematically analyze the $F$-$d$ curves collected for the full range of prepared monolayer densities, even those with high noise-to-signal ratios, we employ a method previously developed for determining elastic moduli from extremely noisy data (O'Bryan et al. 2019, O'Bryan et al. 2024). The method minimizes the contribution of noise to the analysis by comparing the autocorrelation functions of $F$ and $d$, rather than directly fitting $F$-$d$ curves. The analysis method accounts for the contribution of noise, modeling data with a general expression for the measured force, given by

$$F(t) = K\phi(t) + n(t) + F_0,$$

where $\phi(t) = d^p(t)$, $K$ contains the elastic modulus and other constants, $n(t)$ is random noise with zero mean, and $F_0$ is an unknown constant offset in the force data. The power $p$ and the parameters that constitute $K$ depend on the contact mechanics model that is employed to analyze the $F$-$d$ data. Here, we employ the Winkler contact model in which $F \sim d^2$, so $p = 2$. The Winkler model is appropriate to use in contexts where the test material is a very thin slab, having a thickness much less than the contact width; previous results on MDCK monolayers support this choice (Schulze et al. 2017). We will discuss the details of $K$ later. By computing the autocorrelation function of both sides of the $F(t)$ equation, the random noise term is eliminated, as are many cross-terms, yielding a simple relationship between the $F$ and $\phi$ autocorrelation functions, given by

$$C_{FF}(\tau) = K^2 C_{\phi\phi}(\tau) + \beta,$$

where $C_{FF}$ is the $F$ autocorrelation function, $C_{\phi\phi}$ is the $\phi$ autocorrelation function, and $\beta$ is a combination of constants. Thus, $K$ can be determined from a linear regression of $C_{FF}$ versus $C_{\phi\phi}$.

A representative plot of $C_{FF}$ shows the effect of computing the autocorrelation function; a small, sharp drop between $\tau = 0$ and the first time-shift reflects the noise amplitude, while the rest of the curve is dominated by the shape of the $F$ profile. We also note that small and regularly spaced oscillations are visible in $C_{FF}(\tau)$, which likely arise from weak vibrations of the micro-indentation cantilever (Fig. 2b). In contrast to $C_{FF}$, we see no noticeable drop in $C_{\phi\phi}$ around $\tau = 0$, and we see no visible oscillations (Fig. 2c). Thus, measurements of $d$ exhibit much less noise and are much less susceptible to mechanical vibrations than measurements of $F$. Eliminating $\tau$ and parametrically plotting $C_{FF}$ versus $C_{\phi\phi}$, we find a clear linear relationship between the two parameters. The oscillations observed in $C_{FF}(\tau)$ can be seen superimposed on the $C_{FF}$ versus $C_{\phi\phi}$ curves, but do not appear to diminish the linearity of the trends; the slopes of these curves equal $K^2$, which we determine using linear regression (Fig 2d).

**2.3 Relationship between monolayer elastic modulus and cell density**

The clear linearity between $C_{FF}$ and $C_{\phi\phi}$ provides confidence that the Winkler model captures the elasticity of monolayers over short time-scales; the duration from the first point of contact to the final datapoint we include in our analysis is just 2.4 seconds, which is much shorter than that time-scales of active contraction and stiffening of cells in response to applied forces (Fernández et al. 2006). The $F$-$d$ relationship given by the Winkler model is

$$F = \pi E^* \frac{R}{h} d^2,$$

so the relationship between $E^*$ and the best fit values of $K^2$ is given by

$$E^* = \frac{hK}{\pi R},$$

where $R$ is the radius of the hemispherical indenting probe and $h$ is the monolayer thickness. We previously measured the thickness of MDCK monolayers at individual densities in the range of 1400 cells/mm², finding the typical monolayer thickness to be approximately 6 µm. To determine whether $E^*$ varies with cell density, we performed thickness measurements on a series of monolayers seeded at different cell concentrations. Monolayers were prepared in the same way as for indentation tests, described above. Using a 60x oil-immersion objective, we collected laser-scanning confocal z-stacks of fluorescently labeled monolayers (Fig 3a).

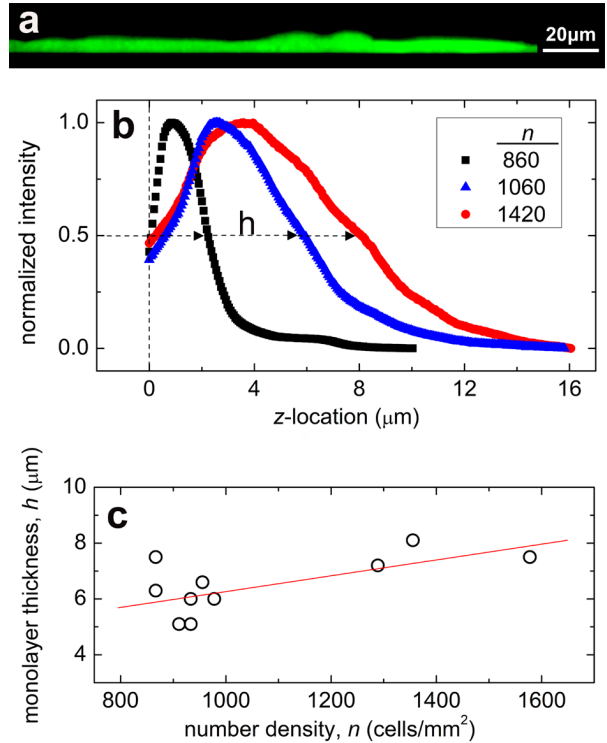

**Figure 3.** Monolayer thickness measurement. (a) X-Z projections of confocal z-stacks show the releatively flat and sporadically bumpy apical surfaces of dyned monolayers, and their extremely flat basal surfaces adhered to glass. (b) The 3D confocal fluorescence intensity profiles are averaged in the X-Y plan to create intensity profiles along the z-axes of monolayers. The location of the half-maximum intensity value is used to determine the monolayer thickness. (c) Monolayer thickness exhibits a weak dependence on cell number density. The best fit line (shown in red) is used to estimate the monolayer thickness for each indentation measurement.

The basal location of each monolayer was determined by manually choosing the first plane in which the monolayer came into focus; using a z-step size of 0.1 µm enables this procedure to be performed with a high degree of certainty, since the basal side of the monolayer is flat. To estimate the location of the apical side of the monolayer, we averaged the 3D fluorescence intensity stack along the $X$ and $Y$ axes, creating an intensity profile along the $Z$-axis, $I(z)$ (Fig. 3b). Normalizing by the maximum intensity of $I(z)$, we estimated the top location of the monolayer by identifying where the normalized intensity drops from its maximum value of 1 to a value of 0.5. Our previous investigation of monolayer thickness showed that this procedure produces the same result as averaging over many local measurements of height (Zehnder Steven M. et al. 2015a). We find that the monolayer thickness varies weakly with cell density. By fitting a line through these datapoints, we create a mapping between cell density and monolayer thickness, which we use to determine $E^*$ from our $F$-$d$ measurements (Fig 3c). Plotting $E^*$ versus $n$, we find that the indentation modulus decreases within increasing cell density in the lower density range, dropping from approximately 30 kPa to about 7kPa before plateauing at high densities (Fig. 4).

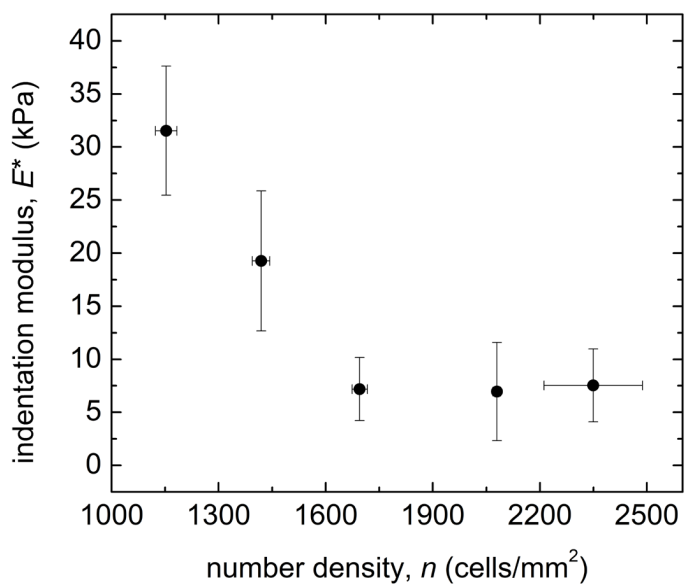

**Figure 4.** Plot of indentation modulus, $E^*$, vs measured number density, $n$. We find a trend of decreasing modulus with increasing density until it reaches a plateau of approximately 7kPa at higher densities (datapoints and intervals correspond to the mean ± standard error averaged over $N$ = 2–9 replicate experiments).

### 3.0 Discussion

In this manuscript we have investigated how the elastic modulus of epithelial monolayers depends on cell packing density. By employing an analysis method that was developed for use with extremely noisy data, we were able to determine the elastic moduli of very soft monolayers that indented to strains of 20% in response to only 10 μN of applied normal load. The results of this analysis method confirmed the choice of the Winkler contact model for interpreting $F$-$d$ curves, as expected from basic geometric considerations and from previous measurements on monolayers prepared at a single density (Schulze et al. 2017). Part of this analysis required measuring the monolayer thickness as a function of density, which we found only to exhibit a weak increase with increasing cell number density.

The most significant finding of this study is that the monolayer's elastic modulus decreases with increasing cell number density in the low-density regime and appears to plateau at the highest densities. Previous work has shown that the elastic modulus of single isolated cells increases with their spread area (Chaudhuri et al. 2015, Nisenholz et al. 2014, Stroka and Aranda-Espinoza 2011), as does the modulus of single cells in monolayers. Thus, our work shows that the monolayer at multi-cellular scales, when tested like an inanimate material slab, exhibits a trend consistent with the previous work characterizing single cells, whether isolated or in monolayer. Each of our indentation experiments simultaneously measure anywhere between a few cells and a few hundred cells throughout an individual *F-d* curve. For example, when using a hemispherical indentation probe having a radius of 1 mm, the contact width at the maximum indentation depth, $d = 2.4$ μm, is approximately $2a = 140$ μm, where $a = \sqrt{2Rd}$. It is noteworthy that this contact width is comparable to the characteristic length-scales of giant cell density fluctuations, previously studied in MDCK layers (Zehnder Steven M. et al. 2015a, Zehnder Steven M. et al. 2015b, Zehnder Steven M. et al. 2016). The work presented here indicates that such giant density fluctuations are likely accompanied by spatially heterogeneous distributions of elastic modulus. In contrast to many passive materials that generally stiffen in non-linear regimes of compression, these results suggest that cells may dynamically soften as they concentrate in the monolayer, tipping toward mechanical instability and leading to giant density fluctuations that are characteristic of active matter (Narayan et al. 2007-7-6).

The potential connection between dynamic fluctuations in cell density and elasticity suggests additional connections to patterns of motion and cell shape within monolayers. For example, as cell density rises in MDCK monolayers, their motion becomes arrested and they exhibit several hallmarks of the glass transition(Angelini et al. 2011). Reevaluating glassiness in dense monolayers while accounting for a density-dependent elastic modulus could reveal new differences between tissue-cell dynamics and collective behaviors in classical molecular or colloidal glasses. Similarly, transitions between fluid-like and solid states with changes in cell shape are predicted by multiple models of monolayers (Bi et al. 2015, Bi et al. 2016). Such a transition has been observed experimentally in lung epithelia where a threshold in cell shape was crossed while unjamming occurred (Park et al. 2015). Cell shape is typically thought to be determined by a balance between cortical tension and the level of cell-cell cohesion. Recent work showed the systematic change in cell shape with changes in cell density (Chisolm et al. 2024). Thus, the softening of monolayers with increased cell density found here is most likely accompanied by changes in cell shape and thus may contribute to transitions in collective patterns of motion within monolayers and transitions between fluid-like and jammed states. Since the states of collective motion in monolayers are so strongly tied to cell shape through the level of cell-cell cohesion, moving forward we plan to investigate how monolayer elasticity at the multi-cellular scale is determined by the combination of cell-cell cohesion and cell density.

## 4.0 Materials and methods

### 4.1 Cell culture and island seeding

Madin Darby canine kidney (MDCK) epithelial cells are cultured in Dulbecco's modified Eagle's medium (DMEM) supplemented with 10% fetal bovine serum (FBS) and 1% penicillin-streptomycin, maintained at 37 °C in a 5% $CO_2$ atmosphere. The cells are grown approximately 70% confluence in six-well plates then harvested using standard trypsinization protocols. The dispersed cells are pelleted with gentle centrifugation and resuspended in 175μL of fresh culture medium. To create monolayer islands, several 2 μL droplets of the cell suspension are deposited in a grid-like pattern near the center of a glass-bottomed petri dish coated with molecular collagen-1 and incubated for 30 minutes to allow cell attachment. 2 mL of fresh culture media is added to the dish, which is then incubated for an additional 12-24 hours before commencing time-lapse imaging. To enable observation of the cells during indentation, the monolayers are treated with a dye solution containing 5-chloromethyl-fluorescein diacetate (CMFDA) and dimethyl sulfoxide (DMSO) in serum-free and antibiotic-free DMEM for 30 minutes. After dye treatment, the cells are washed with and returned to culture media.

To prepare dishes coated with molecular collagen, 35 mm culture dishes having microscope coverslips as their base (Cellvis, product #:D35-20-1.5H) are exposed to 200 Watts UV light for 5 minutes. Molecular collagen at a concentration of 6 mg/mL (Nutragen Type I Collagen Solution, Advanced Biomatrix 5010) is diluted with milli-Q water to a concentration of 0.04 mg/mL. 175 µL of the diluted collagen solution is pipetted onto the coverglass region of the dish and left at room temperature for 30 minutes. The collagen solution is aspirated and the dish is washed with PBS buffer then air dried in a bio safety cabinet (1300 series A2) before cell islands are deposited.

**4.2 Micro-Indentation Instrument**

To perform indentations on MDCK monolayers we designed a micro-indentation system that mounts to the condenser arm of an inverted microscope, in place of the condenser lens turret. This design facilitates the alignment of the indentation probe with the optical axis. The indentation system is made from a capacitance sensor, a piezo-controlled $z$-stage, and a rigid indentation probe mounted onto a calibrated cantilevered flexure. The probe indents the surface of the monolayer as the stage downward, bending the cantilever. Cantilever deflection creates a voltage change measured by the capacitance sensor. We measure the voltage from the indentation onto the monolayers using a voltage input module (National Instruments NI 9220) and compact DAQ (National Instruments cDAQ-9171). This DAQ collects the voltage measurements at 5000 Hz. The linear positioner (Physik Instrumente P-628.1CD) used to move the probe has an 800 µm travel range with 1 nm resolution. Custom-written MATLAB code translates the piezo-stage while collecting position and voltage data. The voltage is converted to a force using the measured cantilever stiffness. Indentation depth is determined by computing the difference between piezo stage position and cantilever deflection.

**5.0 Acknowledgements**

This work was supported by the National Science Foundation under Grant Number 2104429.